\documentclass[10pt]{article}
\usepackage[utf8]{inputenc}
\usepackage[T1]{fontenc}
\usepackage{graphicx}
\usepackage{caption}
\usepackage{subcaption}
\usepackage{amsmath}
\usepackage{setspace}
\usepackage[margin=1.0in]{geometry}
\usepackage{natbib}
\usepackage{longtable}
\usepackage{url}

\title{A Unit Selection Methodology for Music Generation Using Deep Neural Networks}
\author{Mason Bretan\\
\texttt{Georgia Tech}\\
\texttt{Atlanta, GA}\\
\and
Gil Weinberg \\
\texttt{Georgia Tech}\\
\texttt{Atlanta, GA}\\
\and
Larry Heck \\
Google Research \\
\texttt{Mountain View, CA}
}

\date{December 2016}

\begin{document}

\maketitle

\begin{abstract}
Several methods exist for a computer to generate music based on data including Markov chains, recurrent neural networks, recombinancy, and grammars. We explore the use of unit selection and concatenation as a means of generating music using a procedure based on ranking, where, we consider a unit to be a variable length number of measures of music. We first examine whether a unit selection method, that is restricted to a finite size unit library, can be sufficient for encompassing a wide spectrum of music. We do this by developing a deep autoencoder that encodes a musical input and reconstructs the input by selecting from the library. We then describe a generative model that combines a deep structured semantic model (DSSM) with an LSTM to predict the next unit, where units consist of four, two, and one measures of music. We evaluate the generative model using objective metrics including mean rank and accuracy and with a subjective listening test in which expert musicians are asked to complete a forced-choiced ranking task. We compare our model to a note-level generative baseline that consists of a stacked LSTM trained to predict forward by one note.
\end{abstract}

\section {Introduction}
For the last half century researchers and artists have developed many types of algorithmic composition systems. These individuals are driven by the allure of both simulating human aesthetic creativity through computation and tapping into the artistic potential deep-seated in the inhuman characteristics of computers. Some systems may employ rule-based, sampling, or morphing methodologies to create music \citep{papadopoulos1999ai}. We present a method that falls into the class of symbolic generative music systems consisting of data driven models which utilize statistical machine learning.

Within this class of music systems, the most prevalent method is to create a model that learns likely transitions between notes using sequential modeling techniques such as Markov chains or recurrent neural networks \citep{pachet2011markov,franklin2006recurrent}. The learning minimizes note-level perplexity and during generation the models may stochastically or deterministically select the next best note given the preceding note(s).                                                      

In this paper we describe a method to generate monophonic melodic lines based on unit selection. Our work is inspired by a technique that is commonly used in text-to-speech (TTS) systems. The two system design trends found in TTS are statistical parametric and unit selection \citep{zen2009statistical}. In the former, speech is completely reconstructed given a set of parameters. The premise for the latter is that new, intelligible, and natural sounding speech can be synthesized by concatenating smaller audio units that were derived from a preexisting speech signal \citep{hunt1996unit,black1997automatically,conkie2000preselection}. Unlike a parametric system, which reconstructs the signal from the bottom up, the information within a unit is preserved and is directly applied for signal construction. When this idea is applied to music, the generative system can similarly get some of the structure inherent to music ``for free" by pulling from a unit library.

The ability to directly use the music that was previously composed or performed by a human can be a significant advantage when trying to imitate a style or pass a musical Turing test. However, there are also drawbacks to unit selection that the more common note-to-note level generation methods do not need to address. The most obvious drawback is that the output of a unit selection method is restricted to what is available in the unit library. Note-level generation provides maximum flexibility in what can be produced. Ideally, the units in a unit selection method should be small enough such that it is possible to produce a wide spectrum of music, while, remaining large enough to take advantage of the built-in information.

Another challenge with unit selection is that the concatenation process may lead to ``jumps" or ``shifts" in the musical content or style that may sound unnatural and jarring to a listener. Even if the selection process accounts for this, the size of the library must be sufficiently large in order to address many scenarios. Thus, the process of selecting units can equate to a massive number of comparisons among units when the library is very big. Even after pruning this can be a lot of computation. However, this is less of an issue as long as the computing power is available and unit evaluation can be performed in parallel processes.

In this work we explore unit selection as a means of music generation. We first build a deep autoencoder where reconstruction is performed using unit selection. This allows us to make an initial qualitative assessment of the ability of a finite-sized library to reconstruct never before seen music. We then describe a generative method that selects and concatenates units to create new music.

The proposed generation system ranks individual units based on two values: 1) a semantic relevance score between two units and 2) a concatenation cost that describes the distortion at the seams where units connect. The semantic relevance score is determined by using a deep structured semantic model (DSSM) to compute the distance between two units in a compressed embedding space \citep{huang2013learning}. The concatenation cost is derived by first learning the likelihood of a sequence of musical events (such as individual notes) with an LSTM and then using this LSTM to evaluate the likelihood of two consecutive units. We evaluate the model's ability to select the next best unit based on ranking accuracy and mean rank. We use a subjective listening test to evaluate the ``naturalness" and ``likeability" of the musical output produced by versions of the system using units of lengths four, two, and one measures. We additionally compare our unit selection based systems to the more common note-level generative models using an LSTM trained to predict forward by one note.

\section{Related Work}
Many methods for generating music have been proposed. The data-driven statistical methods typically employ n-gram or Markov models \citep{chordia2011predictive,pachet2011markov,wang2014guided,simon2008mysong,collins2016developing}. In these Markov-based approaches note-to-note transitions are modeled (typically bi-gram or tri-gram note models). However, by focusing only on such local temporal dependencies these models fail to take into account the higher level structure and semantics important to music.

Like the Markov approaches, RNN methods that are trained on note-to-note transitions fail to capture higher level semantics and long term dependencies \citep{coca2011generation,boulanger2012modeling,goel2014polyphonic}. However, using an LSTM, Eck demonstrated that some higher level temporal structure can be learned \citep{eck2002finding}. The overall harmonic form of the blues was learned by training the network with various improvisations over the standard blues progression.

We believe these previous efforts have not been successful at creating rich and aesthetically pleasing large scale musical structures that demonstrate an ability to communicate complex musical ideas beyond the note-to-note level. A melody (precomposed or improvised) relies on a hierarchical structure and the higher-levels in this hierarchy are arguably the most important part of generating a melody. Much like in story telling it is the broad ideas that are of the most interest and not necessarily the individual words.

Rule-based grammar methods have been developed to address such hierarchical structure. Though many of these systems' rules are derived using a well-thought out and careful consideration to music theory and perception \citep{lerdahl1992cognitive}, some of them do employ machine learning methods to create the rules. This includes stochastic grammars and constraint based reasoning methods \citep{mccormack1996grammar}. However, grammar based systems are used predominantly from an analysis perspective and do not typically generalize beyond specific scenarios \citep{lerdahl1987generative,papadopoulos1999ai}.

The most closely related work to our proposed unit selection method is David Cope's {\it Experiments in Musical Intelligence}, in which ``recombinancy" is used \citep{cope1999one}. Cope's process of recombinancy first breaks down a musical piece into small segments, labels these segments based on various characteristics, and reorders or ``recombines" them based on a set of musical rules to create a new piece. Though there is no machine learning involved, the underlying process of stitching together preexisting segments is similar to our method. However, we attempt to learn how to connect units based on sequential modeling with an LSTM. Furthermore, our unit labeling is derived from a semantic embedding using a technique developed for ranking tasks in natural language processing (NLP).

Our goal in this research is to examine the potential for unit selection as a means of music generation. Ideally, the method should capture some of the structural hierarchy inherent to music like the grammar based strategies, but be flexible enough so that they generalize as well as the generative note-level models. Challenges include finding a unit length capable of this and developing a selection method that results in both likeable and natural sounding music.

\section{Reconstruction Using Unit Selection}
As a first step towards evaluating the potential for unit selection, we examine how well a melody or a more complex jazz solo can be reconstructed using only the units available in a library. Two things are needed to accomplish this: 1) data to build a unit library and 2) a method for analyzing a melody and identifying the best units to reconstruct it.

Our dataset consists of 4,235 lead sheets from the Wikifonia database containing melodies from genres including (but not limited to) jazz, folk, pop, and classical \citep{simon2008mysong}. In addition, we collected 120 publicly available jazz solo transcriptions from various websites.

\subsection{Design of a Musical DBN Autoencoder}
In order to analyze and reconstruct a melody we trained a deep autoencoder to encode and decode a single measure of music. This means that our unit (in this scenario) is one measure of music. From the dataset there are roughly 170,000 unique measures. Of these, there are roughly 20,000 unique rhythms seen in the measures. We augment the dataset by manipulating pitches through linear shifts (transpositions) and alterations of the intervals between notes resulting in roughly 80 million unique measures.

We augment the dataset by manipulating pitches through linear shifts (transpositions) and alterations of the intervals between notes. We alter the intervals using two methods: 1) adding a constant value to the original intervals and 2) multiplying a constant value to the intervals. Many different constant values are used and the resulting pitches from the new interval values are superimposed on to the measure's original rhythms. The new unit is added to the dataset. We restrict the library to measures with pitches that fall into a five octave range (midi notes 36-92). Each measure is transposed up and down a half step so that all instances within the pitch range are covered. The only manipulation performed on the duration values of notes within a measure is the temporal compression of two consecutive measures into a single measure. This ``double time" representation effectively increases the number of measures, while leaving the inherent rhythmic structure in tact. After all of this manipulation and augmentation there are roughly 80 million unique measures. We use 60\% for training and 40\% for testing our autoencoder.

The first step in the process is feature extraction and creating a vector representation of the unit. Unit selection allows for a lossy representation of the events within a measure. As long as it is possible to rank the units it is not necessary to be able to recreate the exact sequence of notes with the autoencoder. Therefore, we can represent each measure using a bag-of-words (BOW) like feature vector. Our features include:
\begin{enumerate}
\item counts of note tuples \textless pitch$_{1}$, duration$_{1}$\textgreater
\item counts of pitches \textless pitch$_{1}$\textgreater
\item counts of durations \textless duration$_{1}$\textgreater
\item counts of pitch class \textless class$_{1}$\textgreater
\item counts of class and rhythm tuples \textless class$_{1}$, duration$_{1}$\textgreater
\item counts of pitch bigrams \textless pitch$_{1}$, pitch$_{2}$\textgreater
\item counts of duration bigrams \textless duration$_{1}$, duration$_{2}$\textgreater
\item counts of pitch class bigrams \textless class$_{1}$, class$_{1}$\textgreater
\item first note is tied previous measure (1 or 0)
\item last note is tied to next measure (1 or 0)
\end{enumerate}

The pitches are represented using midi pitch values. The pitch class of a note is the note's pitch reduced down to a single octave (12 possible values). We also represent rests using a pitch value equal to negative one. Therefore, no feature vector will consist of only zeros. Instead, if the measure is empty the feature vector will have a value of one at the position representing a whole rest. Because we used data that came from symbolic notation (not performance) the durations can be represented using their rational form (numerator, denominator) where a quarter note would be `1/4.' Finally, we also include beginning and end symbols to indicate whether the note is a first or last note in a measure.

\begin{figure}[h]
  \centering
  \includegraphics[width=0.4\textwidth]{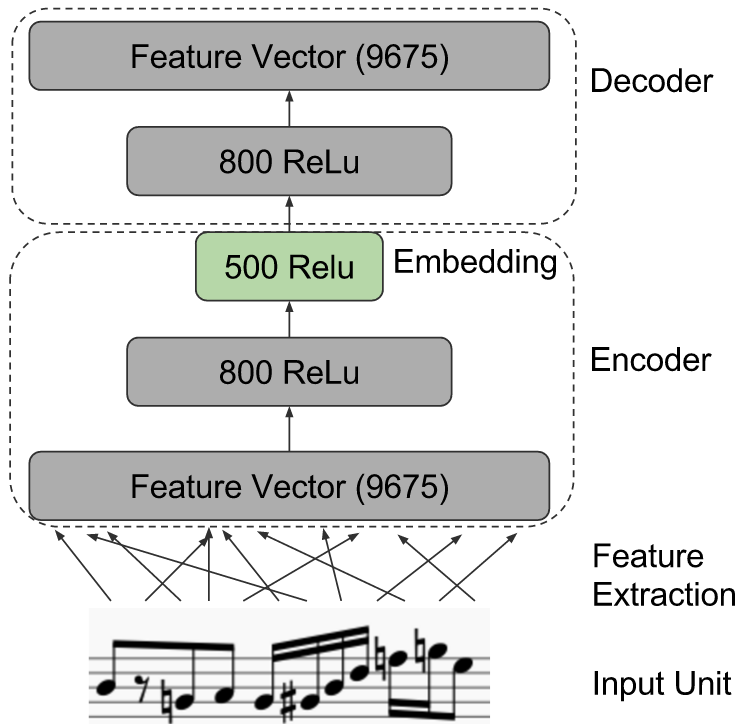}
  \caption{Autoencoder architecture -- The unit is vectorized using a BOW like feature extraction and the autoencoder learns to reconstruct this feature
  vector.}
\end{figure}

The architecture of the autoencoder is depicted in Figure 1. The objective of the decoder is to reconstruct the feature vector and not the actual sequence of notes as depicted in the initial unit of music. Therefore, the entire process involves two types of reconstruction:
\begin{enumerate}
    \item {\bf feature vector reconstruction} - the reconstruction performed and learned by the {\it decoder}.
    \item {\bf music reconstruction} - the process of selecting a unit that best represents the initial input musical unit.
\end{enumerate}
In order for the network to learn the parameters necessary for effective feature vector reconstruction by the decoder, the network uses leaky rectified linear units ($\alpha$ = .001) on each layer and during training minimizes a loss function based on the cosine similarity function

\begin{equation}
\it{sim}(\vec{X}, \vec{Y}) = \frac{\vec{X}^T\cdot \vec{Y}}{|\vec{X} || \vec{Y}|}
\end{equation}
where $\vec{X}$ and $\vec{Y}$ are two equal length vectors. This function serves as the basis for computing the distance between the input vector to the encoder and output vector of the decoder. Negative examples are included through a softmax function

\begin{equation}
\it{\it{P}(\vec{R}|\vec{Q}) = \frac{\exp(\it{sim}(\vec{Q}, \vec{R}))}{\sum_{\vec{d} \epsilon D} \exp(\it{sim}(\vec{Q}, \vec{d}))}}
\end{equation}
where  $\vec{Q}$ is the feature vector derived from the input musical unit, $Q$, and $\vec{R}$ represents the reconstructed feature vector of $Q$. $D$ is the set of five reconstructed feature vectors that includes $\vec{R}$ and four candidate reconstructed feature vectors derived from four randomly selected units in the training set. The network then minimizes the following differentiable loss function using gradient descent

\begin{equation}
\it{-log}\prod_{(Q, R)} P(\vec{R} | \vec{Q})
\end{equation}

A learning rate of 0.005 was used and a dropout of 0.5 was applied to each hidden layer, but not applied to the feature vector. The network was developed using Google's {\it Tensorflow} framework.

\subsection{Music Reconstruction through Selection}
The feature vector used as the input to the autoencoder is a BOW-like representation of the musical unit. This is not a loss-less representation and there is no effective means of converting this representation back into its original symbolic musical form. However, the nature of a unit selection method is such that it is not necessary to reconstruct the original sequence of notes. Instead, a candidate is selected from the library that best depicts the content of the original unit based on some distance metric.

In TTS, this distance metric is referred to as the {\it target cost} and describes the distance between a unit in the database and the target it's supposed to represent \citep{zen2009statistical}. In our musical scenario, the targets are individual measures of music and the distance (or cost) is measured within the embedding space learned by the autoencoder. The unit whose embedding vector shares the highest cosine similarity with the query embedding is chosen as the top candidate to represent a query or target unit. We apply the function 
\begin{equation}
\hat{y} = \arg\max_{y } \it{sim}(x, y)
\end{equation}
where $x$ is the embedding of the input unit and $y$ is the embedding of a unit chosen from the library.

The encoding and selection can be objectively and qualitatively evaluated. For the purposes of this particular musical autoencoder, an effective embedding is one that captures perceptually significant semantic properties and is capable of distinguishing the original unit in the library (low collision rate) despite the reduced dimensionality. In order to assess the second part we can complete a ranking (or sorting) task in which the selection rank (using equation 5) of the truth out of 49 randomly selected units (rank@50) is calculated for each unit in the test set. The collision rate can also be computed by counting the instances in which a particular embedding represents more than one unit. The results are reported in the table below.

\begin{table}
\caption{Results}
\begin{center}
\begin{tabular}{| l | c |}
  \hline			
  mean rank @ 50 & 1.003  \\
  accuracy @ 50 & 99.98 \\
  collision rate per 100k & 91 \\
  \hline  
\end{tabular}
\end{center}
\end{table}

Given the good performance we can make a strong assumption that if an identical unit to the one being encoded exists in the library then the reconstruction process will correctly select it as having the highest similarity. In practice, however, it is probable that such a unit will not exist in the library. The number of ways in which a measure can be filled with notes is insurmountably huge and the millions of measures in the current unit library represent only a tiny fraction of all possibilities. Therefore, in the instances in which an identical unit is unavailable an alternative, though perceptually similar, selection must be chosen.

Autoencoders and embeddings developed for image processing tasks are often qualitatively evaluated by examining the similarity between original and reconstructed images \citep{DBLP:journals/corr/OordKVEGK16}. Likewise, we can assess the selection process by reconstructing never before seen music.

Figure 2 shows the reconstruction of an improvisation (see the related video for audio examples \footnote{\url{https://youtu.be/BbyvbO2F7ug}}). Through these types of reconstructions we are able to see and hear that the unit selection performs well. Also, note that this method of reconstruction utilizes only a target cost and does not include a concatenation cost between measures.

\begin{figure}[h]
  \centering
  \includegraphics[width=0.4\textwidth]{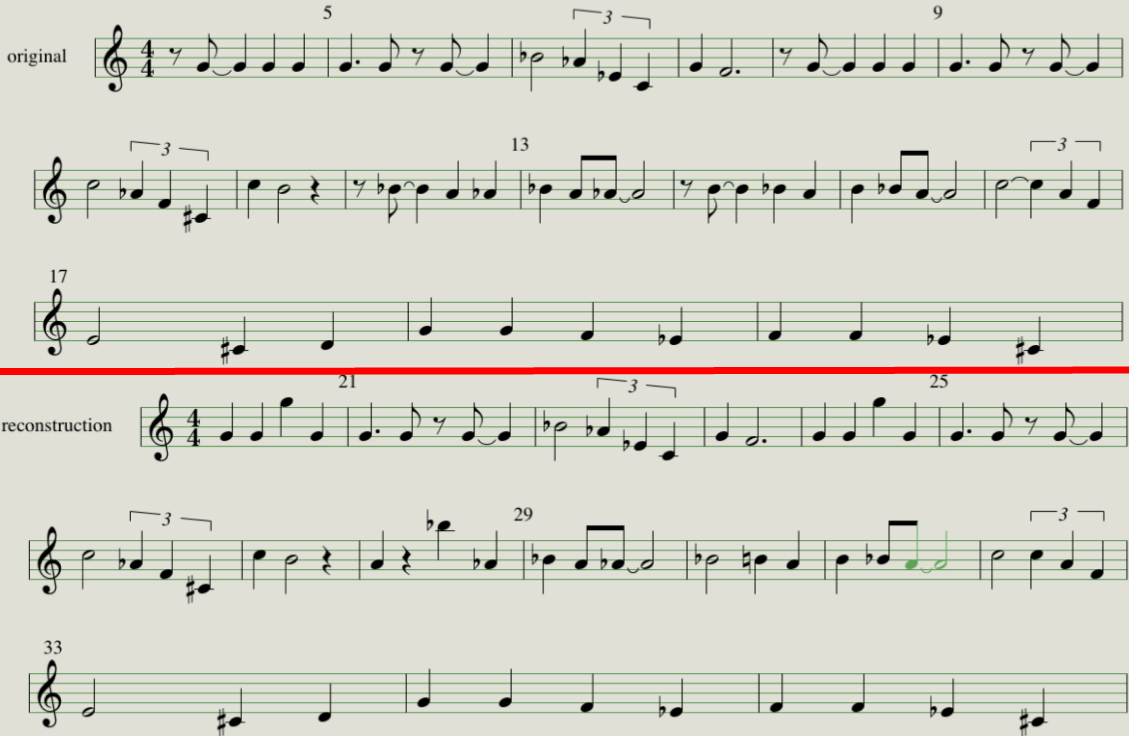}
  \caption{The music on the stave labeled ``reconstruction" (below the line) is the reconstruction (using the encoding and unit selection process) of the music on the stave labeled ``original" (above the line).}
\end{figure}

Another method of qualitative evaluation is to reconstruct from embeddings derived from linear interpolations between two input seeds. The premise is that the reconstruction from the vector representing the weighted sum of the two seed embeddings should result in samples that contain characteristics of both seed units. Figure 3 shows results of reconstruction from three different pairs of units.

\begin{figure}[h]
  \centering
  \includegraphics[width=0.55\textwidth]{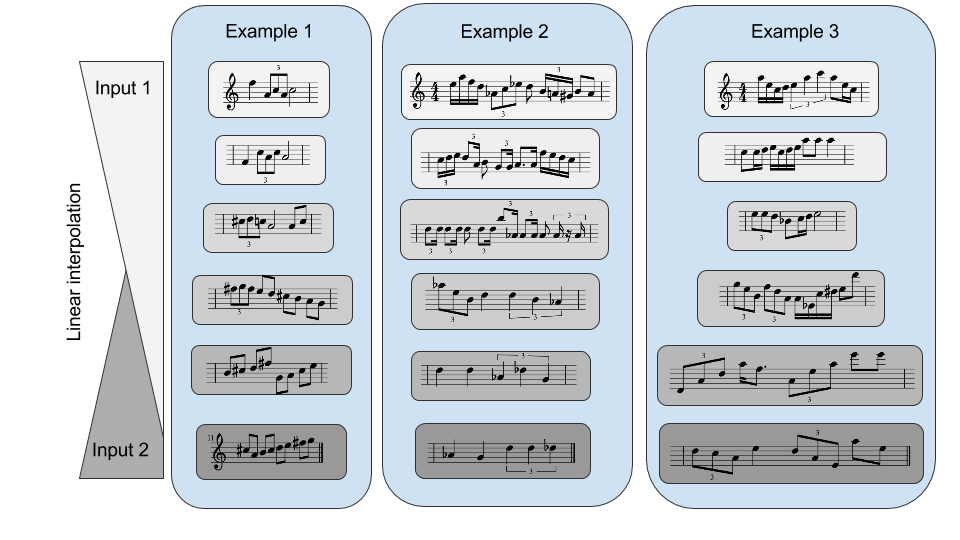}
  \caption{Linear interpolation in the embedding space in which the top and bottom units are used as endpoints in the interpolation. Units are selected based on their cosine similarity to the interpolated embedding vector.}
\end{figure}

\section{Generation using Unit Selection}
In the previous section we demonstrated how unit selection and an autoencoder can be used to transform an existing piece of music through reconstruction and merging processes. The embeddings learned by the autoencoder provide features that are used to select the unit in the library that best represents a given query unit. In this section we explore how unit selection can be used to generate sequences of music using a predictive method. The task of the system is to generate sequences by identifying good candidates in the library to contiguously follow a given unit or sequence of units.

The process for identifying good candidates is based on the assumption that two contiguous units, $(u_{n-1}, u_{n})$, should share characteristics in a higher level musical semantic space (semantic relevance) and the transition between the last and first notes of the first and second units respectively should be likely to occur according to a model (concatenation). This general idea is visually portrayed in Figure 4. We use a DSSM based on BOW-like features to model the semantic relevance between two contiguous units and a note-level LSTM to learn likely note sequences (where a note contains pitch and rhythm information).

\begin{figure}[h]
  \centering
  \includegraphics[width=0.45\textwidth]{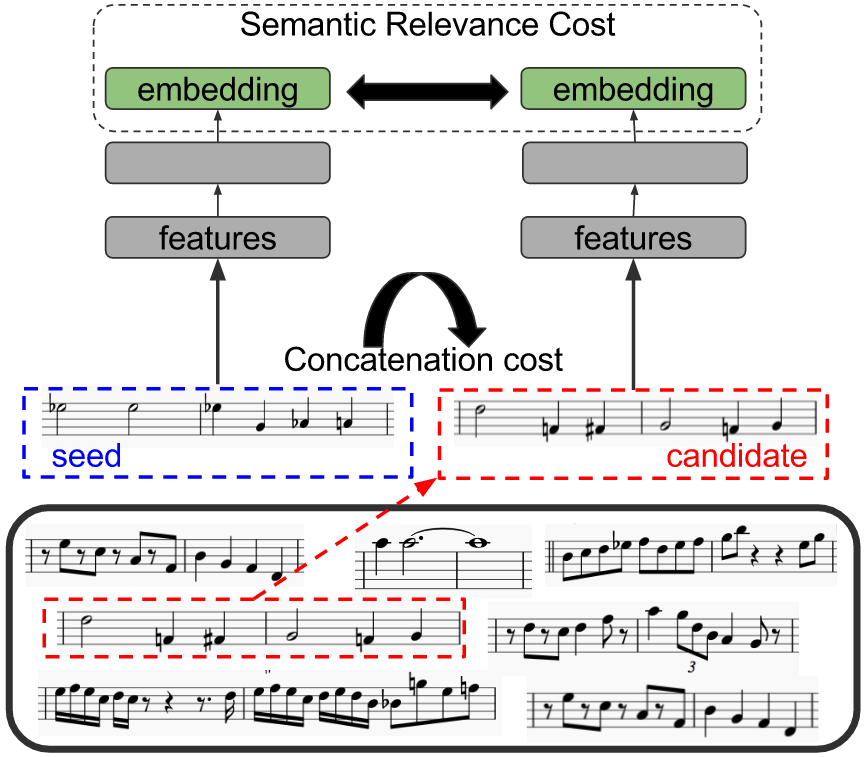}
  \caption{A candidate is picked from the unit library and evaluated based on a concatenation cost that describes the likelihood of the sequence of notes (based on a note-level LSTM) and a semantic relevance cost that describes the relationship between the two units in an embedding space (based on a DSSM).}
\end{figure}

For training these models we use the same dataset described in the previous section. However, in order to ensure that the model learns sequences and relationships that are musically appropriate we can only augment the dataset by transposing the pieces to different keys. Transposing does not compromise the original structure, pitch intervals, or rhythmic information within the data, however, the other transformations do affect these musical attributes and such transformations should not be applied for learning the parameters of these sequential models. However, it is possible to use the original unit library (including augmentations) when selecting units during generation.

\subsection{Semantic Relevance}
In both TTS and the previous musical reconstruction tests a target is provided. For generation tasks, however, the system must predict the next target based on the current sequential and contextual information that is available. In music, even if the content between two contiguous measures or phrases is different, their exist characteristics that suggest the two are not only related, but also likely to be adjacent to one another within the overall context of a musical score. We refer to this likelihood as the ``semantic relevance" between two units.

This measure is obtained from a feature space learned using a DSSM. Though the underlying premise of the DSSM is similar to the DBN autencoder in that the objective is to learn good features in a compressed semantic space, the DSSM features, however, are derived in order to describe the relevance between two different units by specifically maximizing the posterior probability of consecutive units, $P(u_{n}|u_{n-1})$, found in the training data. The same BOW features described in the previous section are used as input to the model. There are two hidden layers and the output layer describes the semantic feature vector used for computing the relevance. Each layer has 128 rectified linear units. The same softmax that was used for the autoencoder for computing loss is used for the DSSM. However, the loss is computed within vectors of the embedding space such that

\begin{equation}
\it{-log}\prod_{(u_{n-1}, u_{n})} P(\vec{u_{n}} | \vec{u_{n-1}})
\end{equation}
where the vectors, $\vec{u_{n}}$ and $\vec{u_{n-1}}$, represent the 128 length embeddings of each unit derived from the parameters of the DSSM. Once the parameters are learned through gradient descent the model can be used to measure the relevance between any two units, $U_{1}$ and $U_{2}$, using cosine similarity $\it{sim}(\vec{U_{1}}, \vec{U_{2}})$ (see Equation 1).

The DSSM provides a meaningful measure between two units, however, it does not describe how to join the units (which one should come first). Similarly, the BOW representation of the input vector does not contain information that is relevant for making decisions regarding sequence. In order to optimally join two units a second measure is necessary.

\subsection{Concatenation Cost}
 By using a unit library made up of original human compositions or improvisations, we can assume that the information within each unit is musically valid. In an attempt to ensure that the music remains valid after combining new units we employ a concatenation cost to describe the quality of the join between two units. This cost requires sequential information at a more fine grained level than the BOW-DSSM can provide.
 
 We use a multi-layer LSTM to learn a note-to-note level model (akin to a character level language model). Each state in the model represents an individual note that is defined by its pitch and duration. This constitutes about a 3,000 note vocabulary. Using a one-hot encoding for the input, the model is trained to predict the next note, $y_{T}$, given a sequence, ${\bf x} = (x_{1}, ..., x_{T})$, of previously seen notes. During training, the output sequence, ${\bf y} = (y_{1}, ..., y_{T})$, of the network is such that $y_{t} = x_{t+1}$. Therefore, the predictive distribution of possible next notes, $\it{Pr}(x_{T+1}|{\bf x})$, is represented in the output vector, $y_T$. We use a sequence length of $T=36$.
 
 The aim of the concatenation cost is to compute a score evaluating the transition between the last note of the unit, $u_{n-1,x_{T}}$, and the first note of the unit, $u_{n, y_{T}}$. By using an LSTM it is possible to include additional context and note dependencies that exist further in the past than $u_{n-1,x_{T}}$. The cost between two units is computed as
 
 \begin{equation}
\it{C}(u_{n-1},u_{n}) = -\frac{1}{J}\sum_{j}^{J}\it{log}\it{Pr}(x_{j}|{\bf x_{j}})
 \end{equation}
 where $J$ is the number of notes in $u_{n}$, $x_{j}$ is the jth note of $u_{n}$, and ${\bf x_{j}}$ is the sequence of notes (with length $T$) immediately before $x_{j}$. Thus, for $j>1$ and $j<T$, ${\bf x_{j}}$ will include notes from $u_{n}$ and $u_{n-1}$ and for $j\geq T$, ${\bf x_{j}}$ will consist of notes entirely from $u_{n}$. In practice, however, the DSSM performs better than the note-level LSTM for predicting the next unit and we found that computing $C$ with $J=1$ provides the best performance. Therefore, the quality of the join is determined using only the first note of the unit in question ($u_{n}$).
 
 The sequence length, $T=36$, was chosen because it is roughly the average number of notes in four measures of music (from our dataset). Unlike the DSSM, which computes distances based on information from a fixed number of measures, the context provided to the LSTM is fixed in the number of notes. This means it may look more or less than four measures into the past. In the scenario in which there is less that 36 notes of available context the sequence is zero padded.

 \subsection{Ranking Units}
 A ranking process that combines the semantic relevance and concatenation cost is used to perform unit selection. Often times in music generation systems the music is not generated deterministically, but instead uses a stochastic process and samples from a distribution that is provided by the model. One reason for this is that note-level Markov chains or LSTMs may get ``stuck" repeating the same note(s). Adding randomness to the procedure helps to prevent this. Here, we describe a deterministic method as this system is not as prone to repetitive behaviors. However, it is simple to apply stochastic decision processes to this system as the variance provided by sampling can be desirable if the goal is to obtain many different musical outputs from a single input seed.
 
The ranking process is performed in four steps:
\begin{enumerate}
    \item Rank all units according to their semantic relevance with an input seed using the feature space learned by the DSSM.
    \item Take the units whose semantic relevance ranks them in the top 5\% and re-rank based on their concatenation cost with the input.
    \item Re-rank the same top 5\% based on their combined semantic relevance and concatenation ranks.
    \item Select the unit with the highest combined rank.
\end{enumerate}

By limiting the combined rank score to using only the top 5\% we are creating a bias towards the semantic relevance. The decision to do this was motivated by findings from pilot listening tests in which it was found that a coherent melodic sequence relies more on the stylistic or semantic relatedness between two units than a smooth transition at the point of connection.

\subsection{Evaluating the model}
The model's ability to choose good units can be evaluated using a ranking test. The task for the model is to predict the next unit given a never before seen four measures of music (from the held out test set). The prediction is made by ranking 50 candidates in which one is the truth and the other 49 are units randomly selected from the database. We repeat the experiments for musical units of different lengths including four, two, and one measures. The results are reported in the table below and they are based on the concatenation cost alone (LSTM), semantic relevance (DSSM), and the combined concatenation and semantic relevance using the selection process described above (DSSM+LSTM).

\begin{table}
\caption{Unit Ranking}
\begin{center}
\begin{tabular}{| l || c | c | c |}
  \hline			
  Model & Unit length & Acc &  Mean Rank \\
   & (measures) & & @50 \\
  \hline
  LSTM & 4 &17.2\% & 14.1 \\
  DSSM & 4 & 33.2\% & 6.9 \\
  DSSM+LSTM & 4 & 36.5\% & 5.9 \\
  \hline
  LSTM  & 2 &16.6\% & 14.8 \\
  DSSM & 2 & 24.4\% & 10.3 \\
  DSSM+LSTM & 2 & 28.0\% & 9.1 \\
\hline
  LSTM & 1 & 16.1\% &15.7  \\
  DSSM & 1 &  19.7\% &16.3  \\
  DSSM+LSTM & 1& 20.6\%  & 13.9 \\
  \hline  
\end{tabular}
\end{center}
\end{table}

\subsection{Discussion}
As stated earlier the primary benefit of unit selection is being able to directly apply previously composed music. The challenge is stitching together units such that the musical results are stylistically appropriate and coherent. Another challenge in building unit selection systems is determining the optimal length of the unit. The goal is to use what has been seen before, yet have flexibility in what the system is capable of generating. The results of the ranking task may indicate that units of four measures have the best performance, yet these results do not provide any information describing the quality of the generated music.

Music inherently has a very high variance (especially when considering multiple genres). It may be that unit selection is too constraining and note-level control is necessary to create likeable music. Conversely, it may be that unit selection is sufficient and given an input sequence there may be multiple candidates within the unit database that are suitable for extending the sequence. In instances in which the ranking did not place the truth with the highest rank, we cannot assume that the selection is ``wrong" because it may still be musically or stylistically valid. Given that the accuracies are not particularly high in the previous task an additional evaluation step is necessary to both evaluate the unit lengths and to confirm that the decisions made in selecting units are musically appropriate. In order to do this a subjective listening test is necessary.

\begin{figure}[h]
  \centering
  \includegraphics[width=.65\textwidth]{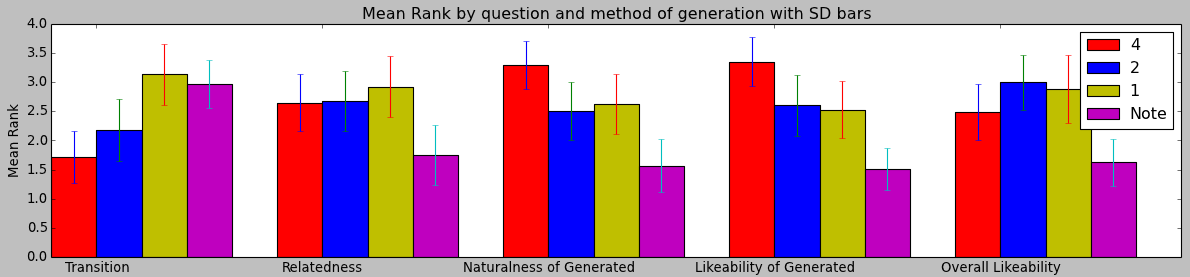}
  \caption{The mean rank and standard deviation for the different music generation systems using units of lengths 4, 2, and 1 measures and note level generation.}
\end{figure}

\begin{figure}[h]
  \centering
  \includegraphics[width=.65\textwidth]{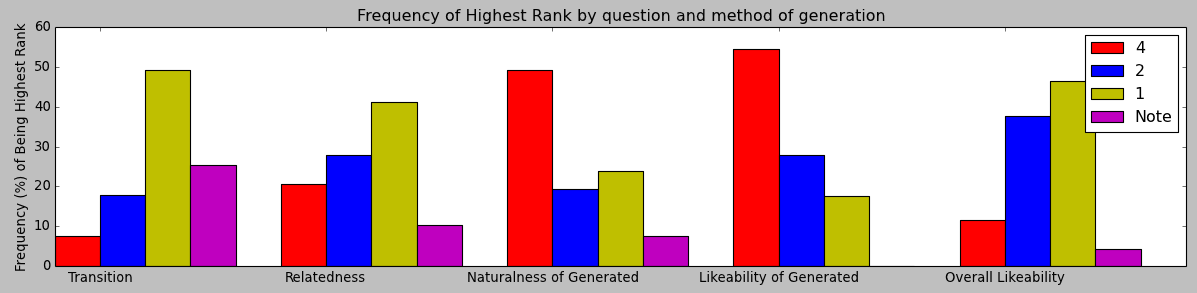}
  \caption{The frequency of being top ranked for the different music generation systems using units of lengths 4, 2, and 1 measures and note level generation. In both Figure 5 and 6 results are reported for each of the five hypotheses: 1) {\bf Transition} -- the naturalness of the transition between the first four measures (input seed) and last four measures (computer generated), 2) {\bf Relatedness} -- the stylistic or semantic relatedness between the first four measures and last four measures, 3) {\bf Naturalness of Generated} -- the naturalness of the last four measures only, 4) {\bf Likeability of Generated} -- the likeability of the last four measures only, and 5) {\bf Overall Likeability} -- the overall likeability of the entire eight measure sequence.}
\end{figure}

\section{Subjective Evaluation}
A subjective listening test was performed. Participants included 32 music experts in which a music expert is defined as an individual that has or is pursuing a higher level degree in music, a professional musician, or a music educator. Four systems were evaluated. Three of the systems employed unit selection using units of four, two, and one measures. The fourth system used the note-level LSTM to generate each note at a time.

The design of the test was inspired by subjective evaluations used by the TTS community. To create a sample each of the four systems was provided with the same input seed (retrieved from the held out dataset) and from this seed each then generated four additional measures of music. This process results in four eight-measure music sequences in which each has the same first four measures. The process was repeated 60 times using random four measure input seeds.

In TTS evaluations participants are asked to rate the quality of the synthesis based on naturalness and intelligibility \citep{stevens2005line}. In music performance systems the quality is typically evaluated using naturalness and likeability \citep{katayose2012evaluating}. For a given listening sample, a participant is asked to listen to four eight-measure sequences (one for each system) and then are asked to rank the candidates within the sample according to questions pertaining to:

\begin{enumerate}
    \item Naturalness of the transition between the first and second four measures.
    \item Stylistic relatedness of the first and second four measures.
    \item Naturalness of the last four measures.
    \item Likeability of the last four measures.
    \item Likeability of the entire eight measures.
\end{enumerate}
Each participant was asked to evaluate 10 samples that were randomly selected from the original 60, thus, all participants listened to music generated by the same four systems, but the actual musical content and order randomly differed from participant to participant. The tests were completed online with an average duration of roughly 80 minutes.

\subsection{Results}
Rank order tests provide ordinal data that emphasize the relative differences among the systems. The average rank was computed across all participants similarly to TTS-MOS tests. The percent of being top ranked was also computed. These are shown in Figures 5 and 6.

In order to test significance the non-parametric Friedman test for repeated measurements was used. The test evaluates the consistency of measurements (ranks) obtained in different ways (audio samples with varying input seeds). The null hypothesis states that random sampling would result in sums of the ranks for each music system similar to what is observed in the experiment. A bonferonni post-hoc correction was used to correct the p-value for the five hypotheses (derived from the itemized question list described earlier).

For each hypothesis the Friedman test resulted in p\textless.05, thus, rejecting the null hypothesis. The sorted ranks for each of the generation system is described in Table 3. 

\begin{table}
\caption{Subjective Ranking}
\begin{center}
\begin{tabular}{| l || c |}
  \hline			
  Variable & Best --\textgreater Worst\\
  \hline
  H1 - Transition Naturalness & 1, N, 2, 4\\
  H2 - Semantic Relatedness &  1, 2, 4, N\\
  H3 - Naturalness of Generated  & 4, 1, 2, N\\
  H4 - Likeability of Generated & 4, 2, 1, N\\
  H5 - Overall Likeability & 2, 1, 4, N \\
  \hline  
\end{tabular}
\end{center}
\end{table}

\subsection{Discussion}
In H3 and H4 the participants were asked to evaluate the quality of the four generated measures alone (disregarding the seed). This means that the sequence resulting from the system that generates units of four measure durations are the unadulterated four measure segments that occurred in the original music. Given there was no computer generation or modification it is not surprising that the four measure system was ranked highest.

The note level generation performed well when it comes to evaluating the naturalness of the transition at the seams between the input seed and computer generated music. However, note level generation does rank highly in the other categories. Our theory is that as the note-level LSTM accumulates error and gets further away from the original input seed the musical quality suffers. This behavior is greatly attenuated in a unit selection method assuming the units are pulled from human compositions.

The results indicate that there exists an optimal unit length that is greater than a single note and less than four measures. This ideal unit length appears to be one or two measures with a bias seemingly favoring one measure. However, to say for certain an additional study is necessary that can better narrow the difference between these two systems.

\section{Conclusion}
We present a method for music generation that utilizes unit selection. The selection process incorporates a score based on the semantic relevance between two units and a score based on the quality of the join at the point of concatenation. Two variables essential to the quality of the system are the breadth and size of the unit database and the unit length. An autoencoder was used to demonstrate the ability to reconstruct never before seen music by picking units out of a database. In the situation that an exact unit is not available the nearest neighbor computed within the embedded vector space is chosen. A subjective listening test was performed in order to evaluate the generated music using different unit durations. Music generated using units of one or two measure durations tended to be ranked higher according to naturalness and likeability than units of four measures or note-level generation.

The system described in this paper generates monophonic melodies and currently does not address situations in which the melodies should conform to a provided harmonic context (chord progression) such as in improvisation. Plans for addressing this are included in future work. Additionally, unit selection may sometimes perform poorly if good units are not available. In such scenarios a hybrid approach that includes unit selection and note-level generation can be useful by allowing the system to take advantage of the structure within each unit whenever appropriate, yet, not restricting the system to the database. Such an approach is also planned for future work.

\selectfont
\bibliography{main}
\bibliographystyle{named}

\end{document}